\documentclass[aps,prd,superscriptaddress,showpacs,preprintnumbers,10pt]{revtex4}
\usepackage[dvips]{graphicx,epsfig,color}
\usepackage{wrapfig,rotating}
\usepackage{amssymb,amsmath,array}

\usepackage{color}

\newcommand{\be}{\begin{equation}}
\newcommand{\ee}{\end{equation}}
\newcommand{\bq}{\begin{eqnarray}}
\newcommand{\eq}{\end{eqnarray}}

\newcommand{\D}{\mathrm{d}}

\def\Vec#1{\mathpalette{\VVec}{#1}}                  
\def\VVec#1#2{\mbox{\boldmath$#1#2$\unboldmath}}

\begin{document}

\title{Transversity distributions from difference asymmetries in semi-inclusive DIS}

\author{V.~Barone}

\affiliation{Di.S.I.T., Universit\`a del Piemonte Orientale, 15121 Alessandria, Italy, \\
and INFN, Sezione di Torino, 10125 Torino, Italy}

\author{F.~Bradamante} 

\author{A.~Bressan} 

\author{A.~Kerbizi}

\author{A.~Martin}

 \author{A.~Moretti} 

\author{J.~Matousek} 

\author{G.~Sbrizzai}

\affiliation{Dipartimento di Fisica, Universit\`a di Trieste, \\
and INFN, Sezione di Trieste, 34127 Trieste, Italy}

\begin{abstract}
In recent years information on the transversity distribution $h_1$ has been obtained
combining the Collins asymmetry results from semi-inclusive deep inelastic scattering
(SIDIS) data on transversely polarized
nucleon targets with the information on the fragmentation function of a transversely polarized quark
from the asymmetries measured in $e^+e^-$ annihilation into hadrons. 
An alternative method was proposed
long time ago,  which does not require the $e^+e^-$ data, but allows one to get ratios of the $u$ 
and $d$ quark transversity distributions from the SIDIS data alone. 
The method utilizes the ratio of the
difference of the Collins asymmetries of positively and negatively charged hadrons produced on
transversely polarized proton and deuteron targets. We have applied this method 
to the COMPASS proton and deuteron data, and extracted the ratio $h_1^d/h_1^u$. 
The results are compared to those obtained in a previous point--by--point extraction based
both on SIDIS and $e^+e^-$ data. 

\end{abstract}


\maketitle

\section{Introduction}

Much interest has been dedicated in the past twenty years to the transversity distribution. Usually
called $h_1$, it is a leading-twist parton distribution function (PDF) which describes the transverse
polarization of quarks inside a transversely polarized nucleon (for reviews,
see~\cite{Barone:2010zz, Aidala:2012mv, Avakian:2016rst}).

Being chirally-odd, transversity cannot be measured in deep inelastic scattering (DIS). 
Over the
last decade single-spin asymmetries clearly related to the transversity distribution function
have been measured in semi-inclusive deep inelastic scattering (SIDIS) on transversely polarized
nucleons, namely in DIS processes in which at least one hadron of the current jet is detected. 
In these processes the cross-section exhibits a spin-dependent azimuthal modulation which can be
expressed in terms of a convolution of the transversity PDF and a fragmentation function (FF) which
is also chiral-odd, thus guaranteeing the cross-section to be chirally-even. 
Two observables have been studied in so far. 
The first one is single hadron spin asymmetry, namely the amplitude of the
target-spin dependent azimuthal modulation of each of the produced hadrons. 
The second one is the amplitude of the target spin-dependent
azimuthal modulation of the plane defined by any two of the oppositely charged hadrons 
produced in the same SIDIS event.
In the case of transversely polarized proton targets
clear non zero azimuthal modulations have been measured for both observables 
by the HERMES~\cite{Airapetian:2010ds,Airapetian:2008sk} and the
COMPASS~\cite{Adolph:2012sn,Adolph:2014fjw} Collaborations, assessing beyond any doubt that both the
transversity PDFs and the single hadron and the dihadron FF are not zero. 
Corresponding measurements
on a transversely polarized deuteron target by the COMPASS Collaboration~\cite{Ageev:2006da,Adolph:2012nw}
 gave asymmetries compatible with zero, which have been interpreted as evidence of
cancellation between $h_1^u$ and $h_1^d$. 

The underlying physics of these processes~\cite{Adolph:2015zwe,Kerbizi:2018qpp}
is the left-right asymmetry in the hadronization of a
transversely polarized quark, where left and right are relative to the plane defined by the quark
direction of motion and its transverse spin. Such asymmetry is encoded, in the first case, in the
so-called Collins FF $H_1^\perp$~\cite{Collins:1992kk}, and, in the second case 
in the dihadron FF $H^\sphericalangle_1$ ~\cite{Collins:1994kq,Radici:2001na}. 
Independent evidence that both the Collins
function and the dihadron FF are different from zero came from the measurements of azimuthal
asymmetries in hadron inclusive production in $e^+e^-$ annihilation by the 
Belle~\cite{Seidl:2008xc,Vossen:2011fk},
the BaBar~\cite{TheBABAR:2013yha} and the BESIII~\cite{Ablikim:2015pta}
Collaborations.

Combining the SIDIS data and the $e^+e^- \rightarrow \mathrm{hadrons}$ measurements, first
extractions of both the transversity functions and of the two transversely polarized quark FFs have
been possible~\cite{Anselmino:2015sxa,Kang:2015msa}. 
In all those works, in order to solve the
convolution over the transverse momenta between the transversity PDF and the FF which appears in the
cross-section, some parametrization for both $h_1$ and for the FFs had to be assumed. 
An
exception is the recent extraction of transversity~\cite{Martin:2014wua} which has been performed
point-by-point directly from the COMPASS SIDIS and the Belle $e^+e^- \rightarrow \mathrm{hadrons}$
data, without using any parametrization for the collinear variables. 

An alternative way to measure transversity from the Collins asymmetries alone
is via the so-called ``difference asymmetries'',  
which allow extracting combinations of the $u$ and $d$ quark transversity without knowing the Collins FF.
This method was proposed a long time ago~\cite{Frankfurt:1989wq,Christova:2000nz,Sissakian:2006vz}
to access the helicity PDFs, and has been used by the SMC Collaboration~\cite{Adeva:1995yi}. It was
also included in the COMPASS proposal~\cite{Baum:1996yv}, to measure both longitudinal
and transverse spin asymmetries. At that time it looked particularly interesting, since
the Collins FF was completely unknown. Later on it has been used to measure the helicity PDFs in
COMPASS~\cite{Alekseev:2007vi},  and recently it has been proposed again in the context of the Sivers,
Boer-Mulders and transversity distributions~\cite{Christova:2015waz}.  In the present work 
the 
difference asymmetries are used for the first time to access transversity 
using the COMPASS measurements of the Collins asymmetries 
on $p$~\cite{Adolph:2012sn} and $d$ targets~\cite{Ageev:2006da}. 

The paper is organized as
follows. In Section~\ref{sec2} we describe the formalism and the procedure. Section~\ref{sec3} is
dedicated to the Monte Carlo studies. The results are discussed in Section~\ref{sec4}.

\section{Cross sections and difference asymmetries}
\label{sec2}

In this paper we will extract the asymmetries of differences from the Collins asymmetries measured
by the COMPASS Collaboration impinging a 160 GeV/c momentum muon beam either on a transversely
polarized deuteron ($^6$LiD) target or a transversely polarized proton (NH$_3$) target. The data we
have considered were taken in the years 2002--2004~\cite{Ageev:2006da} and 
2010~\cite{Adolph:2012sn}.

In order to ensure the DIS regime, only events with photon virtuality $Q^2>1$ (GeV/c)$^2$, fractional
energy of the virtual photon $0.1<y<0.9$, and mass of the hadronic final state system $W>5$ GeV/c$^2$
were considered in the data analysis. The charged hadrons were required to have at least 0.1 GeV/c
transverse momentum $P_T$ with respect to the virtual photon direction and a fraction of the available
energy $z>0.2$. 
All the details of the event selection and of the analysis can be found in \cite{Ageev:2006da,Adolph:2012sn}.
The published data are binned in $x$, the target nucleon momentum fraction carried by the struck
quark, in $z$ and in $P_T$. 
In our analysis we will only consider the asymmetry data binned in $x$,
in 9 bins, from 0.003 to 0.7.

In the following, for simplicity we will write explicitly only the Collins part of the SIDIS
transverse spin dependent cross-section, and consider charged pions, even if, at the
end, we will use the results for charged hadrons assuming they are all pions, as it was done, 
for instance,
in \cite{Martin:2014wua}.  
The SIDIS cross section can be written as
\begin{eqnarray}
\sigma_t^{\pm} (\Phi_C) = \sigma_{0,t}^{\pm} + f P_T D_{NN} \sigma_{C,t}^{\pm} \sin \Phi_C
\label{eq:first}
\end{eqnarray}
where $\Phi_C$ is the Collins angle, $f$ is the target dilution factor, $P_T$ is the nucleon polarization,
and $D_{NN}$ is the mean transverse-spin-transfer coefficient not included in $\sigma_{C}$ to simplify the
expressions used in the following.  
Only the deuteron (or hydrogen) nuclei in the targets
were polarized, and the target dilution factor $f$ is given by the ratio of the absorption 
cross-sections on the deuteron (or proton) to that of all nuclei in the target.
The signs $\pm$ refer to the pion charge and $t=p,\ d$ is the
target type. The Collins angle $\Phi_C=\phi_h+\phi_S-\pi$ is the sum of the azimuthal angles $\phi_h$
of the hadron transverse momentum and of the spin direction $\phi_S$ of the target nucleon with
respect to the lepton scattering plane, in a reference system in which the $z$ axis is the virtual
photon direction.

The Collins asymmetry is defined as 
\begin{eqnarray}
A^{\pm}_{C,t} = \frac{\sigma^{\pm}_{C,t}}{\sigma^{\pm}_{0,t}}
\end{eqnarray}

In terms of the ordinary PDFs and FFs 
the unpolarized part of the cross-sections in eq.~(\ref{eq:first}) can be written as (omitting 
a kinematic factor that cancels out when taking the ratios 
of cross sections):
\begin{eqnarray}
\sigma^{+}_{0,p} &\sim& x \left [ (4 f_1^u + f_1^{\bar{d}})D_{1, {\rm fav}} 
+ (4 f_1^{\bar{u}} + f_1^d)D_{1, {\rm unf}}
                 + (f_1^{s} + f_1^{\bar{s}})D_{1,s} \right ]  \label{unpol1} \\
\sigma^{-}_{0,p} &\sim& x \left [(4 f_1^u + f_1^{\bar{d}})D_{1, {\rm unf}} 
+ (4 f_1^{\bar{u}} + f_1^{d})D_{1, {\rm fav}}
                 + (f_1^{s} + f_1^{\bar{s}})D_{1, s} \right ] \label{unpol2} \\
\sigma^{+}_{0,d} &\sim& x \left [(f_1^u + f_1^d)(4D_{1, {\rm fav}} + D_{1, {\rm unf}})
                 + (f_1^{\bar{u}} + f_1^{\bar{d}})(D_{1, {\rm fav}} + 4 D_{1, {\rm unf}})
                 + 2(f_1^{s} + f_1^{\bar{s}})D_{1, s} \right ]  \label{unpol3} \\
\sigma^{-}_{0,d} &\sim& x \left [(f_1^u + f_1^d)(D_{1, {\rm fav}} + 4D_{1, {\rm unf}})
                 + (f_1^{\bar{u}} + f_1^{\bar{d}})(4D_{1, {\rm fav}} + D_{1, {\rm unf}})
                 + 2(f_1^{s} + f_1^{\bar{s}})D_{1, s} \right ]  
\label{unpol4}
\end{eqnarray}
where $D_{1, {\rm fav}}$ ($D_{1, {\rm unf}}$) is the favored (unfavored) unpolarized FF, 
$D_{1,s}$ is the 
strange sea unpolarized FF, and $f_1^q$ are the unpolarized PDFs.  

Following
\cite{Martin:2014wua},  the corresponding spin--dependent cross sections are obtained by replacing
$f_1^q$ with the transversity PDFs $h_1^q$ and the FFs $D_1$ with the ``half moments'' 
of the Collins function, $H_{1}^{\perp (1/2)}$, defined as 
\be
H_1^{\perp (1/2)}(z, Q^2) \equiv \int \D^2  \Vec p_T 
\, \frac{p_T}{z M_h} \,  
\, H_1^{\perp}(z, p_T^2, Q^2) \,.    
\ee
Thus we have: 
\begin{eqnarray}
\sigma^{+}_{C,p} &\sim& 
x \left [(4 h_1^u + h_1^{\bar{d}})H_{1,{\rm fav}}^{\perp (1/2)} 
+ (4 h_1^{\bar{u}} + h_1^d)H_{1,{\rm unf}}^{\perp (1/2)} \right ] \label{pol1} \\
\sigma^{-}_{C,p} &\sim& 
x \left [(4 h_1^u + h_1^{\bar{d}})H_{1,{\rm unf}}^{\perp (1/2)} + (4 h_1^{\bar{u}} 
+ h_1^{d})H_{1,{\rm fav}}^{\perp (1/2)} \right ] \label{pol2} \\
\sigma^{+}_{C,d} &\sim& 
x \left [(h_1^u + h_1^d)(4H_{1,{\rm fav}}^{\perp (1/2)} + H_{1,{\rm unf}}^{\perp (1/2)})
                 + (h_1^{\bar{u}} + h_1^{\bar{d}})(H_{1,{\rm fav}}^{\perp (1/2)} 
+ 4 H_{1,{\rm unf}}^{\perp (1/2)}) \right ]                   \label{pol3} \\
\sigma^{-}_{C,d} &\sim& 
x \left [ (h_1^u + h_1^d)(H_{1,{\rm fav}}^{\perp (1/2)} + 4H_{1,{\rm unf}}^{\perp (1/2)})
                 + (h_1^{\bar{u}} + h_1^{\bar{d}})(4H_{1,{\rm fav}}^{\perp (1/2)} 
+ H_{1,{\rm unf}}^{\perp (1/2)})\right ] 
\label{pol4}
\end{eqnarray}
where we have assumed $H_{1, s}^{\perp (1/2)}=0$.

We now define the difference asymmetries as 
\begin{eqnarray}
A_{D,t}=\frac{\sigma^{+}_{C,t}-\sigma^{-}_{C,t}}{\sigma^{+}_{0,t}+\sigma^{-}_{0,t}}.
\label{diff1}
\end{eqnarray}
In \cite{Christova:2000nz} an alternative definition was proposed, namely
\begin{eqnarray}
A'_{D,t}=\frac{\sigma^{+}_{C,t}-\sigma^{-}_{C,t}}{\sigma^{+}_{0,t}-\sigma^{-}_{0,t}}.
\label{diff2}
\end{eqnarray}
As we will see, the two definitions turn out to give the same results. 
For the sake of simplicity, our discussion in the following 
we be centered on the definition (\ref{diff1}), but we will also briefly 
summarize the results obtained with  
eq.~(\ref{diff2}).  

Writing explicitly the asymmetries one gets:
\begin{eqnarray}
A_{D,p}&=& \frac{1}{9}
  \frac{H_{1,{\rm fav}}^{\perp (1/2)}-H_{1,{\rm unf}}^{\perp (1/2)}}{\sigma^{+}_{0,p}+\sigma^{-}_{0,p}}
  (4 h_1^{u_v}-h_1^{d_v}) \label{diff_expl1} \\
A_{D,d}&=& \frac{1}{3}
  \frac{H_{1,{\rm fav}}^{\perp (1/2)}-H_{1,{\rm unf}}^{\perp (1/2)}}{\sigma^{+}_{0,d}+\sigma^{-}_{0,d}}
  (h_1^{u_v}+h_1^{d_v}) . \label{diff_expl1bis} 
\end{eqnarray}

When taking the ratios of the asymmetries on deuteron and proton, the Collins FFs cancel out: 
\begin{eqnarray}
\frac{A_{D,d}}{A_{D,p}}= 3  
\left [ \frac{(4 f_1^u + 4 f_1^{\bar u} + f_1^d + f_1^{\bar{d}}) (D_{1, {\rm fav}} + D_{1, {\rm unf}}) 
                 + 2 (f_1^{s} + f_1^{\bar{s}}) D_{1,s}}
{5 (f_1^u + f_1^d + f_1^{\bar u} + f_1^{\bar d})(D_{1, {\rm fav}} + D_{1, {\rm unf}})
                 + 4(f_1^{s} + f_1^{\bar{s}})D_{1, s}} \right ]
 \frac{h_1^{u_v}+h_1^{d_v}}{4 h_1^{u_v}-h_1^{d_v}}, 
\label{ratio1}
\end{eqnarray}
and the only unknowns are the transversity PDFs.  Thus, by measuring $A_D$ on $p$ and $d$, one
obtains the ratio $h_1^{d_v}/h_1^{u_v}$ in terms of known quantities.

In order to determine $A_{D,t}$, one should in principle fit the quantity
\begin{eqnarray}
\sigma^{D}_t(\Phi_C) = (\sigma^{+}_{0,t}-\sigma^{-}_{0,t})+ f P_T D_{NN}
(\sigma^{+}_{C,t}-\sigma^{-}_{C,t}) \sin \Phi_C
\end{eqnarray}
and extract the amplitude of the $\sin \Phi_C$ modulation.  Since usually the acceptances
for positively and negatively charged particles are not the
same, one should correct the number of events for the acceptance before taking the
differences, and treat carefully the statistical errors.  

The measurements are much simpler if the $\Phi_C$ acceptance for positively charged particles is
equal to that for negatively charged ones.  In this case it is not necessary to evaluate the difference
asymmetries from the amplitude of the modulation, as it is possible to get them from the 
measured Collins asymmetries.  One has in fact
\begin{eqnarray}
A_{D,t}&=&\frac{\sigma^{+}_{0,t}}{\sigma^{+}_{0,t}+\sigma^{-}_{0,t}} A^+_{C,t} -
       \frac{\sigma^{-}_{0,t}}{\sigma^{+}_{0,t}+\sigma^{-}_{0,t}} A^-_{C,t},  
\label{eq:dacoll1}
\end{eqnarray}
where the ratios of the cross sections are known.
In order to apply this procedure, extensive Monte Carlo studies have been performed. They are
described in the next Section.

Notice that if one uses instead the definition (\ref{diff2}), 
the ratio of the difference asymmetries has the form 
\begin{equation}
\frac{A'_{D,d}}{A'_{D,p}}= 
  \frac{4 f_1^{u_v}-f_1^{d_v}}{f_1^{u_v}+f_1^{d_v}}
  \frac{h_1^{u_v}+h_1^{d_v}}{4 h_1^{u_v}-h_1^{d_v}} 
\label{ratio2}
\end{equation} 
and the equivalent of eq.~(\ref{eq:dacoll1}) is
\begin{equation}
A'_{D,t}=\frac{\sigma^{+}_{0,t}}{\sigma^{+}_{0,t}-\sigma^{-}_{0,t}} A^+_{C,t} -
       \frac{\sigma^{-}_{0,t}}{\sigma^{+}_{0,t}-\sigma^{-}_{0,t}} A^-_{C,t}.  
\label{eq:dacoll2}
\end{equation}

\section{Monte Carlo studies}
\label{sec3}

The acceptance of the COMPASS spectrometer for positively charged and negatively charged hadrons
have been investigated with Monte Carlo simulations.
In the case of the deuteron data, collected in the years 2002-2004,
this work was a prerequisite to the extraction of the
$\sin \phi_h$, $\cos \phi_h$ and $\cos 2\phi_h$ modulations~~\cite{Adolph:2014pwc}
which are expected in the  unpolarized SIDIS cross-section.
Within the statistical errors, the  acceptance turned out to be essentially the
same for positive and negative hadrons. 
In 2010, when the proton data were collected, the spectrometer was
substantially different from the one utilized for the deuteron data taking were taken, 
thus the whole work had to be repeated.
To this end we have
used a full Monte Carlo chain using LEPTO~\cite{Ingelman:1996mq} as event generator and TGEANT, a
GEANT4~\cite{Asai:2015xno} based program, for the simulation of the particle interaction with the
COMPASS apparatus and the detector response. The Monte Carlo events have been reconstructed with
the COMPASS package
CORAL~\cite{Abbon:2007pq}, and analyzed to extract the acceptances and the acceptance ratios. 
The same
kinematic selections  used for the analysis of the real data have been applied on the generated 
variables and on the reconstructed ones.
While integrating over the other kinematical variables,  
the acceptances ${\mathcal A} (x) = { N^{\rm rec} ( x)}/{N^{\rm gen}(x^{\rm gen})}$ 
have been obtained by taking the ratio of the
reconstructed and generated events counted in every bin using respectively 
the generated $x^{\rm gen}$ and the reconstructed
$x^{\rm rec}$ values. In this way also the smearing due to the experimental
resolution is accounted for.

\begin{figure}
\begin{center}
 \includegraphics[width=\textwidth]{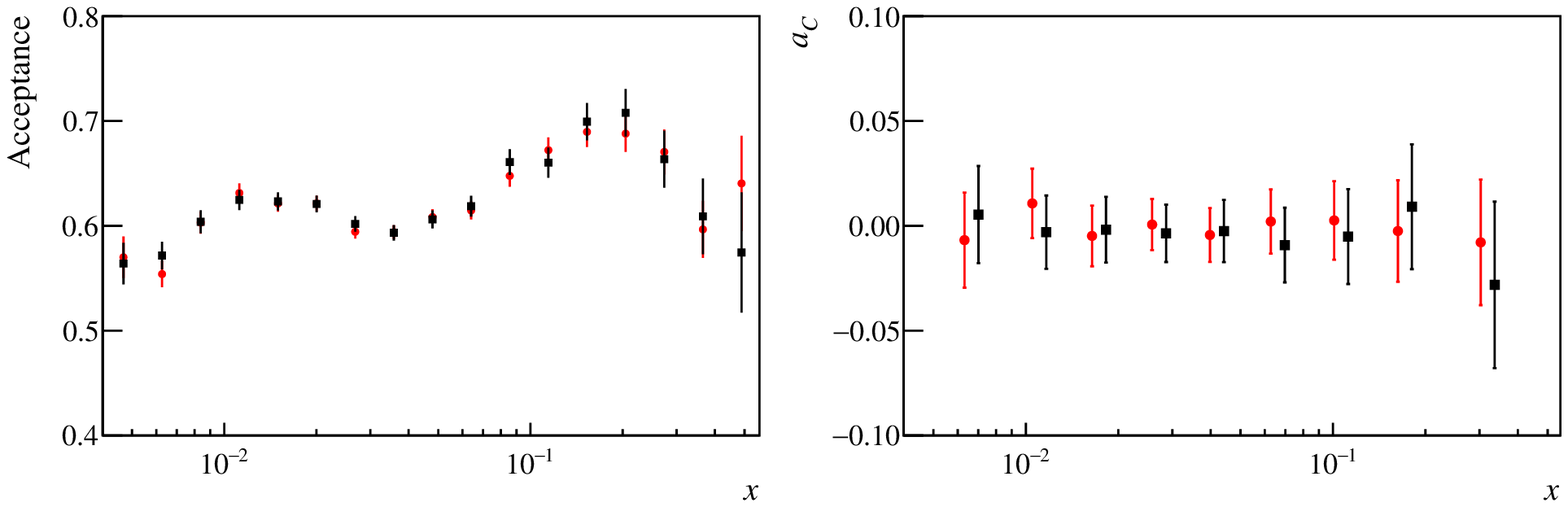}
 \caption{Left:
   Experimental acceptance for positively charged (red points) and negatively charged (black points)
   hadrons as a function of $x$. Right:   Amplitude of the $a_C$ 
modulation in the azimuthal acceptance as a function of
   $x$  for positively (red points) and negatively (black squared)
   charged hadrons.
\label{fig:mcres}}
\end{center}
  \end{figure}

The acceptances, which include both the geometrical acceptance of the apparatus and the
reconstruction efficiency, are shown in Fig.~\ref{fig:mcres} (left). The acceptances for positively
(red points) and negatively (black points) charged hadrons are in good agreement and the small
differences are compatible with the statistical fluctuations.
Their ratios is constant over the full $x$ range of the measurement, with
an average value of $1.003 \pm 0.006$.

A possible Collins modulation in the acceptance was also studied,
separately for positively and negatively charged
hadrons, by  fitting in each $x$ bin the $\Phi_C$ distribution  with 
a function $c \, (1 +a_C \sin(\Phi_C))$. 
The results for $a_C$
are shown in Fig.~\ref{fig:mcres} (right). 
The amplitudes of the modulation are compatible with zero
over the full $x$ range for both positive and negative hadrons. 
This result stays true also
when repeating the procedure for  the ratio of the  acceptances. 
 
\section{Results}
\label{sec4}

\begin{figure}
\begin{center}
\includegraphics[width=0.6\textwidth]{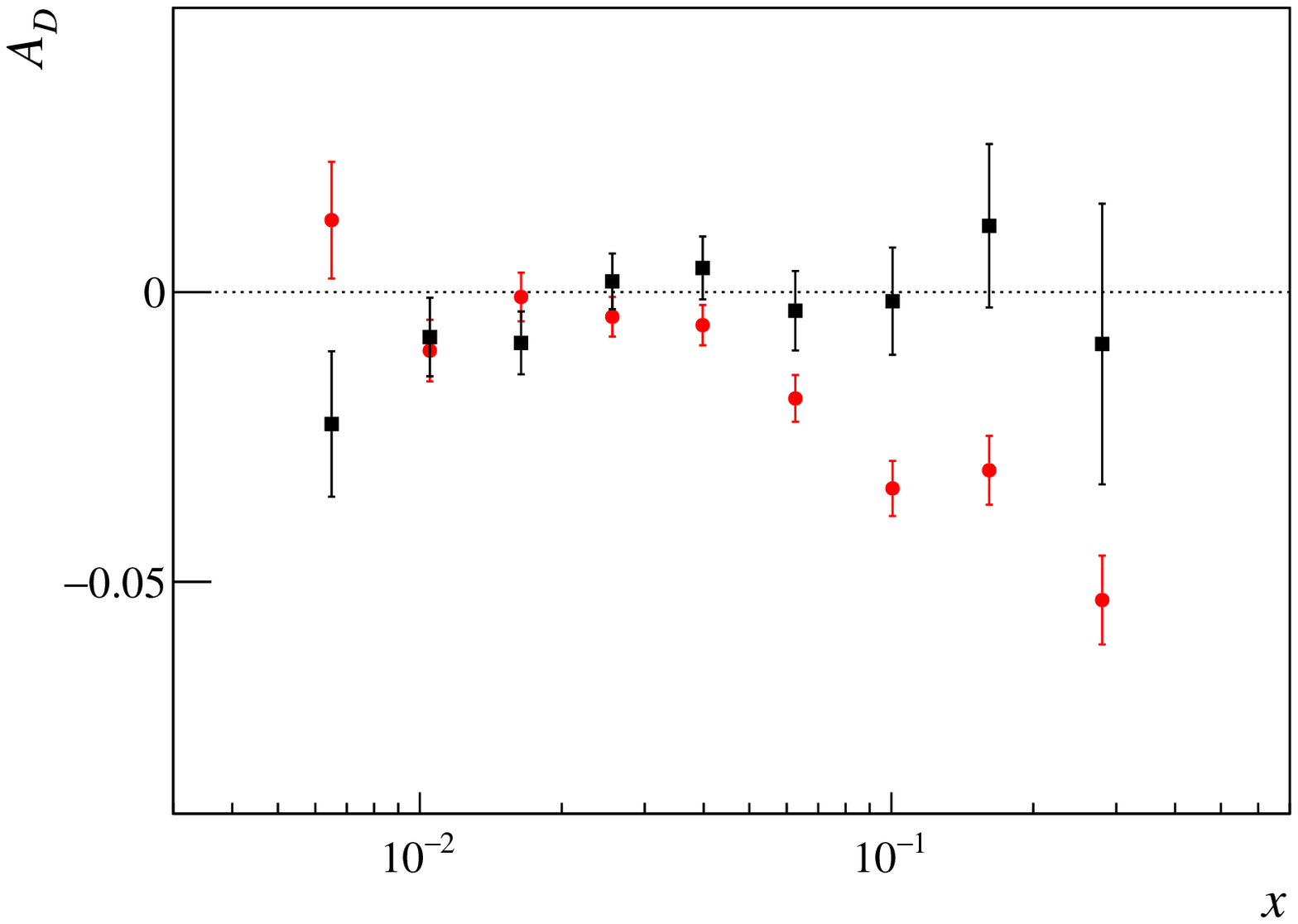}
 \caption{Difference asymmetries $A_{D,p}$ (red points) and $A_{D,d}$ (black points) as function of
   $x$.
\label{fig:adiff_pd}}
\end{center}
  \end{figure}

On the basis of the Monte Carlo results, the difference asymmetries have been calculated using
eq.~(\ref{eq:dacoll1}) with the Collins asymmetries from the 2010 COMPASS data.  
Actually,
since $\sigma^\pm_{0,t} \sim N^\pm_t$ and $\mathrm{var} (A^\pm_{C,t}) \sim 1/ N^\pm_t$, where
$N^\pm_t$ is the total number of hadrons which has been used to extract the Collins asymmetries,
in a given $x$ bin,
eq.~(\ref{eq:dacoll1}) can be rewritten as:
\begin{eqnarray}
A_{D,t}=\frac{\mathrm{var} (A^{-}_{C,t})}{\mathrm{var} (A^{+}_{C,t})+\mathrm{var} (A^{-}_{C,t})}
A^+_{C,t} - \frac{\mathrm{var} (A^{+}_{C,t})}{\mathrm{var} (A^{+}_{C,t})+\mathrm{var} (A^{-}_{C,t})}
A^-_{C,t} \label{eq:dacoll1bis}  
\end{eqnarray}
The calculation of the difference asymmetries  can thus be performed using the published COMPASS data for
$A^\pm_{C,t}$ and their statistical uncertainties~\cite{Adolph:2012sn}. 
An interesting remark is that $A_{D,t}$ is equal to the weighted mean of the Collins asymmetries
for positive and negative hadrons, after changing sign to $A^{-}_{C,t}$.
The results for proton and deuteron are shown in Fig. \ref{fig:adiff_pd}.  

\begin{figure}
\begin{center}
 \includegraphics[width=0.6\textwidth]{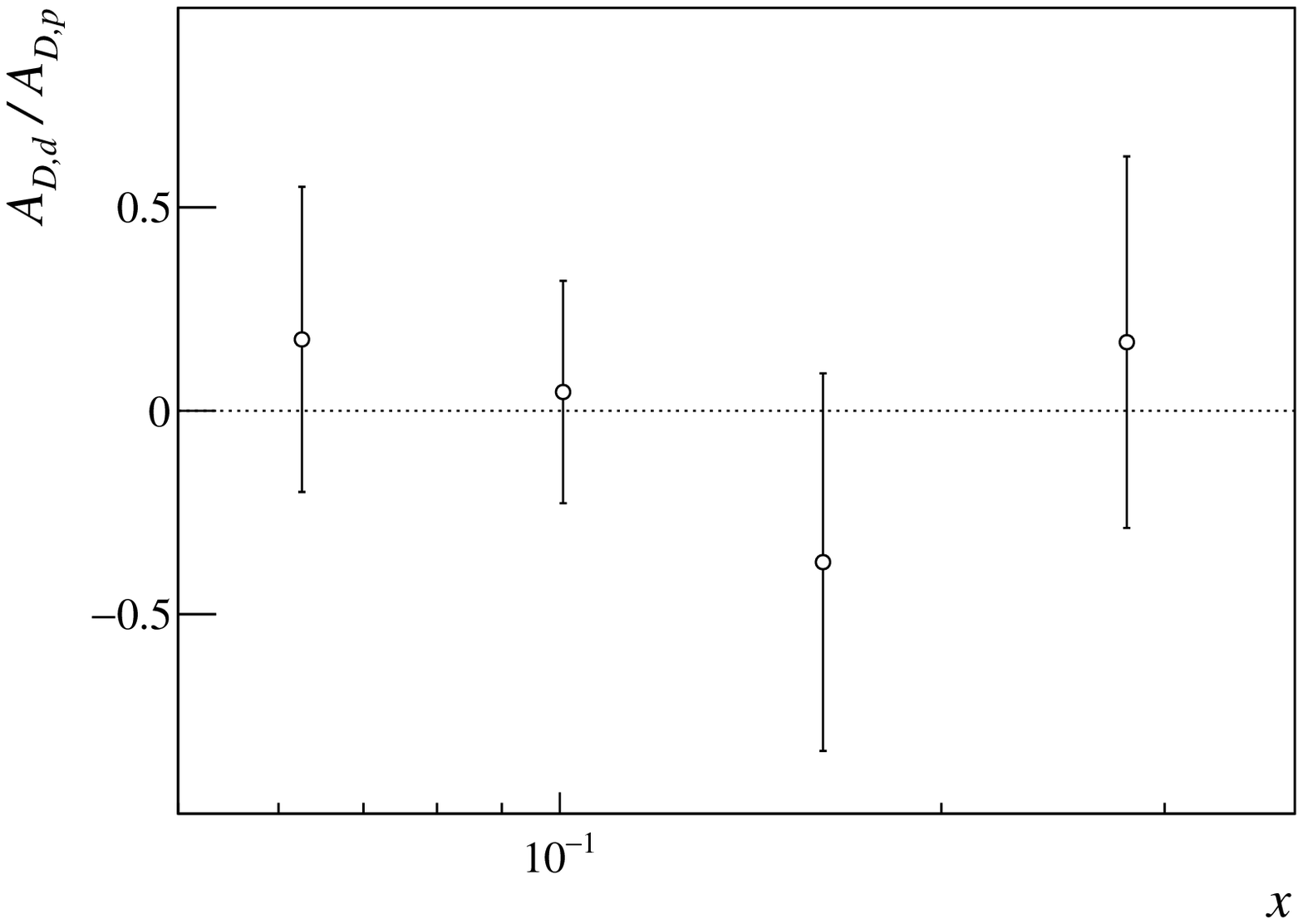}
 \caption{Ratio $A_{D,d}/A_{D,p}$ of the difference asymmetries on deuteron and  on proton as
   function of $x$.  Here and in the next
   figure the ratio in the missing $x$ bins has values out of scale with very large statistical errors.
\label{fig:radiff}}
\end{center}
  \end{figure}

The ratio $A_{D,d}/A_{D,p}$ is shown in  Fig.~\ref{fig:radiff}. 
Only the four points at larger $x$ are plotted in the figure. The points
at smaller $x$ have much too large uncertainties since the proton asymmetries in that region 
are compatible with zero.

From the ratios $A_{D,d}/A_{D,p}$ the quantities $(h_1^{u_v}+h_1^{d_v})/(4h_1^{u_v}-h_1^{d_v})$ 
have been extracted using eq.~(\ref{ratio1}) and standard parametrizations and tables for the unpolarized
PDFs \cite{Lai:1999wy} and FFs \cite{deFlorian:2007aj}. 

\begin{figure}
\begin{center}
\includegraphics[width=0.6\textwidth]{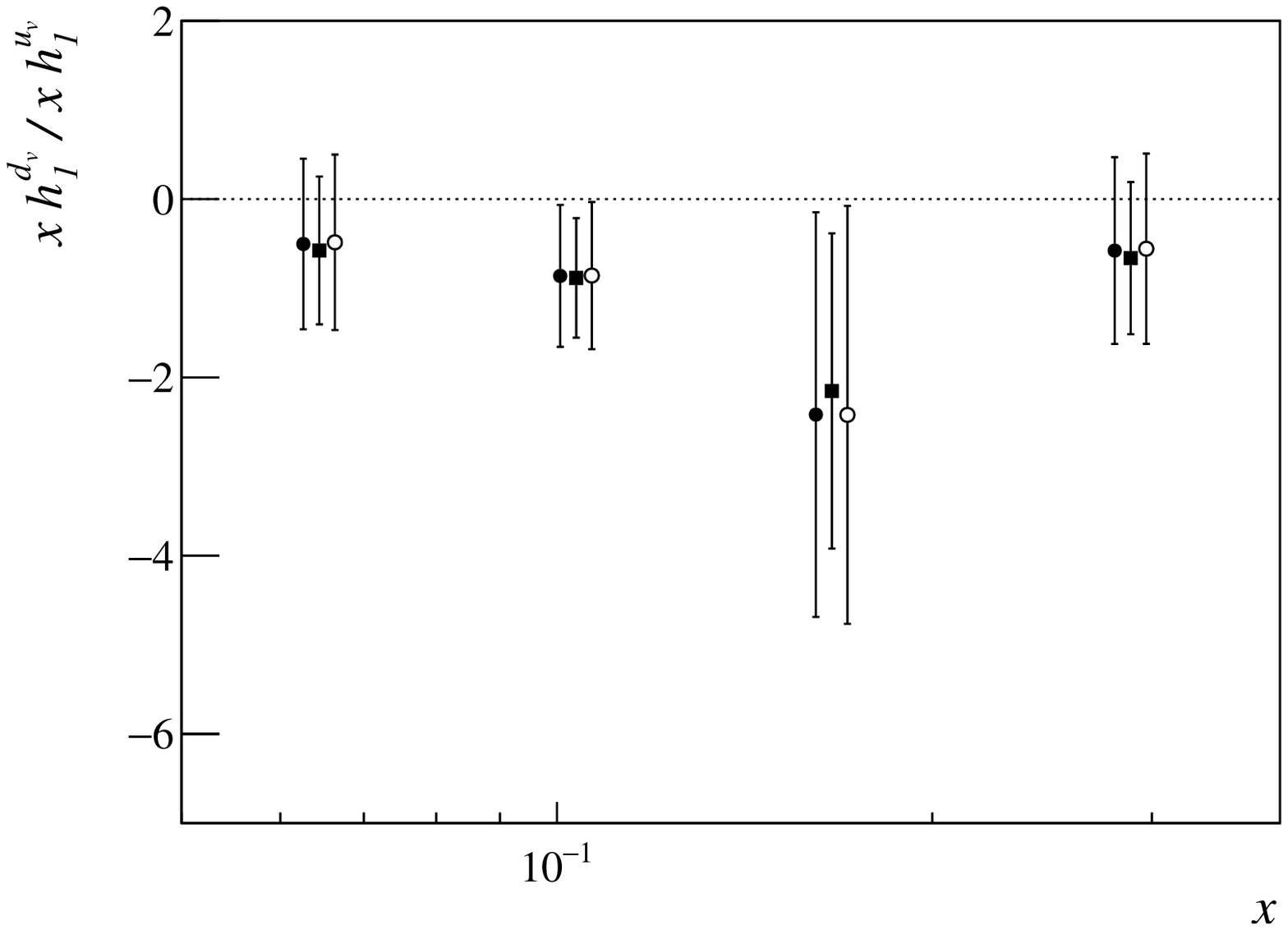}
 \caption{Ratio $h_1^{d_v}/h_1^{u_v}$ from the asymmetries $A_D$ 
(closed circles), from the asymmetries $A'_D$ (closed squares) and from \cite{Martin:2014wua}
   (open circles).
\label{fig:hratio_bbm}}
\end{center}
  \end{figure}

Finally, from the quantities
$(h_1^{u_v}+h_1^{d_v})/(4h_1^{u_v}-h_1^{d_v})$ the ratios $h_1^{d_v}/h_1^{u_v}$ are determined.  
They are shown as closed circles in Fig.~\ref{fig:hratio_bbm}. 
Again, the values in the first five $x$ bins have very large
uncertainties, are compatible with zero and are not plotted in the figure. 
At larger $x$ the values
are negative, in agreement with previous extractions. 
The same procedure has been carried through starting from the difference asymmetries
$A'_{D,t}$ and using eq. (\ref{ratio2}), getting essentially the same values and the same
statistical uncertainties, which are shown as closed squares in  Fig. \ref{fig:hratio_bbm}.
In the same figure we also compare our results with the
values of $h_1^{d_v}/h_1^{u_v}$ calculated from the transversity values obtained
in \cite{Martin:2014wua} (open
circles).  In the evaluation of the uncertainty of the ratio $h_1^{d_v}/h_1^{u_v}$ from
\cite{Martin:2014wua}, proper account has been taken of the correlations between the extracted
values of $h_1^{u_v}$ and $h_1^{d_v}$, and use has been made of the correlation coefficients as
evaluated in~\cite{addendum3}. The results of the three determinations are in very good agreement, 
but some reduction (up to $\sim 20\%$) of the uncertainties can be observed in the ratios obtained
in the present work from the difference asymmetries. 

\section{Conclusion}

We have determined for the first time the transverse-spin difference asymmetries of positively and
negatively charged hadrons using the SIDIS $p$ and $d$ COMPASS data. 
Thanks to the good COMPASS spectrometer acceptance it could be easily obtained from the 
measured Collins asymmetries.
From the ratio between the difference asymmetries on deuterons and on protons we have 
extracted the quantity $h_1^{d_v}/h_1^{u_v}$, the ratio between the valence $d$-quark and $u$-quark 
transversity PDF. 

At small $x$ the difference asymmetries on the protons are compatible with zero, thus the 
statistical uncertainty on the ratio $h_1^{d_v}/h_1^{u_v}$
is too large and no useful information is provided by the present analysis. 
On the other hand, for larger $x$ ($x \leq 0.05$) the extracted ratio  $h_1^{d_v}/h_1^{u_v}$
has  negative sign and is in very good agreement with the results of a previous point by point 
extraction.

The method we applied is interesting and simple, and does not require any knowledge of the Collins
fragmentation functions. Hence it strengthens the validity of the methods utilized so far to
extract the transversity distributions, based on a combined analysis of SIDIS and 
$e^+e^-$
data, and can be used as a useful cross-check for more elaborated extractions.

\begin{acknowledgments}
This work has been possible thanks to the project FRA2015, supported by the Universit\`a degli Studi
di Trieste.  V.B. has been partially supported by 
``Fondi di Ricerca Locale ex--60\%'' of the University of Piemonte 
Orientale. We would like to thank E.~Christova and E.~Leader
for  interesting discussions. We are grateful to the COMPASS Collaboration for the use of the
Monte Carlo chain.  
\end{acknowledgments}


\begin{thebibliography}{10}

\bibitem{Barone:2010zz}
V. Barone, F. Bradamante and A. Martin,
\newblock Prog. Part. Nucl. Phys. {\bf 65}, 267 (2010), arXiv:1011.0909.

\bibitem{Aidala:2012mv}
C.A. Aidala et~al.,
\newblock Rev. Mod. Phys. {\bf 85}, 655 (2013), arXiv:1209.2803.

\bibitem{Avakian:2016rst}
H. Avakian, A. Bressan and M. Contalbrigo,
\newblock Eur. Phys. J. {\bf A52}, 150 (2016),
\newblock [Erratum: Eur. Phys. J. {\bf A52}, 165 (2016)].

\bibitem{Airapetian:2010ds}
A. Airapetian et~al. (HERMES Collaboration),
\newblock Phys. Lett. B {\bf 693}, 11 (2010), arXiv:1006.4221.

\bibitem{Airapetian:2008sk}
A. Airapetian et~al. (HERMES Collaboration),
\newblock J. High Energy Phys. 06 (2008) 017, arXiv:0803.2367.

\bibitem{Adolph:2012sn}
C. Adolph et~al. (COMPASS Collaboration),
\newblock Phys. Lett. B {\bf 717}, 376 (2012), arXiv:1205.5121.

\bibitem{Adolph:2014fjw}
C. Adolph et~al. (COMPASS Collaboration),
\newblock Phys. Lett. B {\bf 736}, 124 (2014), arXiv:1401.7873.

\bibitem{Ageev:2006da}
E.S. Ageev et~al. (COMPASS Collaboration),
\newblock Nucl. Phys. {\bf B765}, 31 (2007), arXiv:hep-ex/0610068.

\bibitem{Adolph:2012nw}
C. Adolph et~al. (COMPASS Collaboration),
\newblock Phys.Lett. B {\bf 713}, 10 (2012), arXiv:1202.6150.

\bibitem{Adolph:2015zwe}
C. Adolph et~al. (COMPASS Collaboration),
\newblock Phys. Lett. B {\bf 753}, 406 (2016), arXiv:1507.07593.

\bibitem{Kerbizi:2018qpp}
A. Kerbizi et~al.,
\newblock Phys. Rev. D {\bf 97}, 074010 (2018), arXiv:1802.00962.

\bibitem{Collins:1992kk}
J.C. Collins,
\newblock Nucl. Phys. {\bf B396}, 161 (1993), arXiv:hep-ph/9208213.

\bibitem{Collins:1994kq}
J.C. Collins, S.F. Heppelmann and G.A. Ladinsky,
\newblock Nucl. Phys. {\bf B420}, 565 (1994), arXiv:hep-ph/9305309.

\bibitem{Radici:2001na}
M. Radici, R. Jakob and A. Bianconi,
\newblock Phys. Rev. D {\bf 65}, 074031 (2002), arXiv:hep-ph/0110252.

\bibitem{Seidl:2008xc}
R. Seidl et~al. (Belle Collaboration),
\newblock Phys. Rev. D {\bf 78}, 032011 (2008), arXiv:0805.2975,
\newblock [Erratum: Phys. Rev. D {\bf 86}, 039905 (2012)].

\bibitem{Vossen:2011fk}
A. Vossen et~al. (Belle Collaboration),
\newblock Phys.Rev.Lett. {\bf 107}, 072004 (2011), arXiv:1104.2425.

\bibitem{TheBABAR:2013yha}
J.P. Lees et~al. (BaBar Collaboration),
\newblock Phys. Rev. D {\bf 90}, 052003 (2014), arXiv:1309.5278.

\bibitem{Ablikim:2015pta}
M. Ablikim et~al. (BESIII Collaboration),
\newblock Phys. Rev. Lett. {\bf 116}, 042001 (2016), arXiv:1507.06824.

\bibitem{Anselmino:2015sxa}
M. Anselmino et~al.,
\newblock Phys. Rev. D {\bf 92}, 114023 (2015), arXiv:1510.05389,
\newblock and references therein.

\bibitem{Kang:2015msa}
Z.B. Kang et~al.,
\newblock Phys. Rev. D {\bf 93}, 014009 (2016), arXiv:1505.05589,
\newblock and references therein.

\bibitem{Martin:2014wua}
A. Martin, F. Bradamante and V. Barone,
\newblock Phys. Rev. D {\bf 91}, 014034 (2015), arXiv:1412.5946.

\bibitem{Frankfurt:1989wq}
L.L. Frankfurt et~al.,
\newblock Phys. Lett. B {\bf 230}, 141 (1989).

\bibitem{Christova:2000nz}
E. Christova and E. Leader,
\newblock Nucl. Phys. {\bf B607}, 369 (2001), arXiv:hep-ph/0007303.

\bibitem{Sissakian:2006vz}
A.N. Sissakian, O.{\relax Yu}. Shevchenko and O.N. Ivanov,
\newblock Phys. Rev. D {\bf 73}, 094026 (2006), arXiv:hep-ph/0603236.

\bibitem{Adeva:1995yi}
B. Adeva et~al. (Spin Muon Collaboration),
\newblock Phys. Lett. B {\bf 369}, 93 (1996).

\bibitem{Baum:1996yv}
G. Baum et~al. (COMPASS Collaboration),
\newblock {COMPASS: A Proposal for a Common Muon and Proton Apparatus for
  Structure and Spectroscopy},
\newblock CERN-SPSLC-96-14, CERN-SPSLC-P-297, 1996.

\bibitem{Alekseev:2007vi}
M. Alekseev et~al. (COMPASS Collaboration),
\newblock Phys. Lett. B {\bf 660}, 458 (2008), arXiv:0707.4077.

\bibitem{Christova:2015waz}
E. Christova and E. Leader,
\newblock J. Phys. Conf. Ser. 678, 012012 (2016), arXiv:1512.01404.

\bibitem{Adolph:2014pwc}
C. Adolph et~al. (COMPASS Collaboration),
\newblock Nucl. Phys. {\bf B886}, 1046 (2014), arXiv:1401.6284.

\bibitem{Ingelman:1996mq}
G. Ingelman, A. Edin and J. Rathsman,
\newblock Comput. Phys. Commun. {\bf 101}, 108 (1997), arXiv:hep-ph/9605286.

\bibitem{Asai:2015xno}
M. Asai et~al. (Geant4 Collaboration),
\newblock Annals Nucl. Energy {\bf 82}, 19 (2015).

\bibitem{Abbon:2007pq}
P. Abbon et~al. (COMPASS Collaboration),
\newblock Nucl.\ Instrum.\ Meth. {\bf A577}, 455 (2007), arXiv:hep-ex/0703049.

\bibitem{Lai:1999wy}
H.L. Lai et~al. (CTEQ Collaboration),
\newblock Eur. Phys. J. C {\bf 12}, 375 (2000), arXiv:hep-ph/9903282.

\bibitem{deFlorian:2007aj}
D. de~Florian, R. Sassot and M. Stratmann,
\newblock Phys. Rev. D {\bf 75}, 114010 (2007), arXiv:hep-ph/0703242.

\bibitem{addendum3}
K. Augsten et~al. (COMPASS Collaboration),
\newblock {Addendum to the COMPASS-II Proposal, d-Quark Transversity and Proton
  Radius},
\newblock CERN-SPSC-2017-034 SPSC-P-340-ADD-1, 2018.

\end{thebibliography}

\end{document}